\documentclass{PoS}

\title{The TeV emitter structure in LS~5039}

\ShortTitle{The TeV emitter structure in LS~5039}

\author{Valent\'i Bosch-Ramon\\
        Max Planck Institut f\"ur Kernphysik, Saupfercheckweg 1, Heidelberg 69117, Germany\\
        E-mail: \email{vbosch@mpi-hd.mpg.de}}

\author{Dmitry Khangulyan\\
        Max Planck Institut f\"ur Kernphysik, Saupfercheckweg 1, Heidelberg 69117, Germany\\
        E-mail: \email{dmitry.khangulyan@mpi-hd.mpg.de}}
	
\author{Felix Aharonian\\
        Dublin Institute for Advanced Studies, Dublin, Ireland; 
	Max Planck Institut f\"ur Kernphysik, Saupfercheckweg 1, Heidelberg 69117, Germany\\
        E-mail: \email{felix.aharonian@mpi-hd.mpg.de;felix.aharonian@dias.ie}}
	
\abstract{LS~5039 is an X-ray binary detected at very high energies. Along the orbit, there is a significant 
detection even 
during the superior conjunction of
the compact object, when very large gamma-ray opacities are expected. Electromagnetic cascades, 
which may make the system more transparent
to gamma-rays, are hardly efficient for reasonable magnetic fields in the massive
star surroundings. A jet-like flow could transport energy to
regions where the photon-photon absorption is much lower and the TeV radiation is
not so severely absorbed. Otherwise, in the standard pulsar scenario for LS~5039, 
the emitter would be located between the star and the compact object, which would imply 
the violation of the observational constraints at X-rays.}

\FullConference{VII Microquasar Workshop: Microquasars and Beyond\\
		 September 1-5 2008\\
		 Foca, Izmir, Turkey}

\begin{document}

\section{Introduction}

LS~5039 is a high-mass X-ray binary located at distances of 2.5~kpc \cite{Casares et
al. 2005} with extended non-thermal radio emission \cite{Paredes et al.
2000}. This source has been detected in the very high-energy (VHE) range all
along the orbit and also during the superior conjunction of the compact object
\cite{Aharonian et al. 2005,Aharonian et al. 2006}, when the photon-photon absorption is expected to be
the strongest \cite{Dubus 2006a}.
LS~5039 is formed by an O star and a compact object of unknown nature \cite{Casares et al. 2005}. Although thought to be a
microquasar after the discovery of the extended radio emission (e.g. \cite{Paredes et al. 2000,Paredes et al. 2006}), the non detection of X-ray
accretion features led \cite{Martocchia et al. 2005,Dubus 2006b} to suggest the presence of a non-accreting pulsar in the
system. 

In this work, we point out that, under reasonable values for the ambient magnetic field, 
whatever its nature the TeV emitter should be located in
the borders of the binary system at least around the superior conjunction of the compact object. Otherwise, the
electron-positron pairs created via photon-photon absorption in the stellar photon field would radiate overcoming the
observed X-ray emission levels. 

\section{The TeV emitter in LS~5039}

\begin{figure}
\caption{Absorbed luminosity $\times 1/4\pi d_{\odot}^2$, 
depending on the location of the emitter in the binary system. The contour line limits the
region to the left, in which the emitter, if located there, would overcome the observed X-ray fluxes if the synchrotron 
emission is the dominant radiation channel.}
\includegraphics[height=.65\textheight]{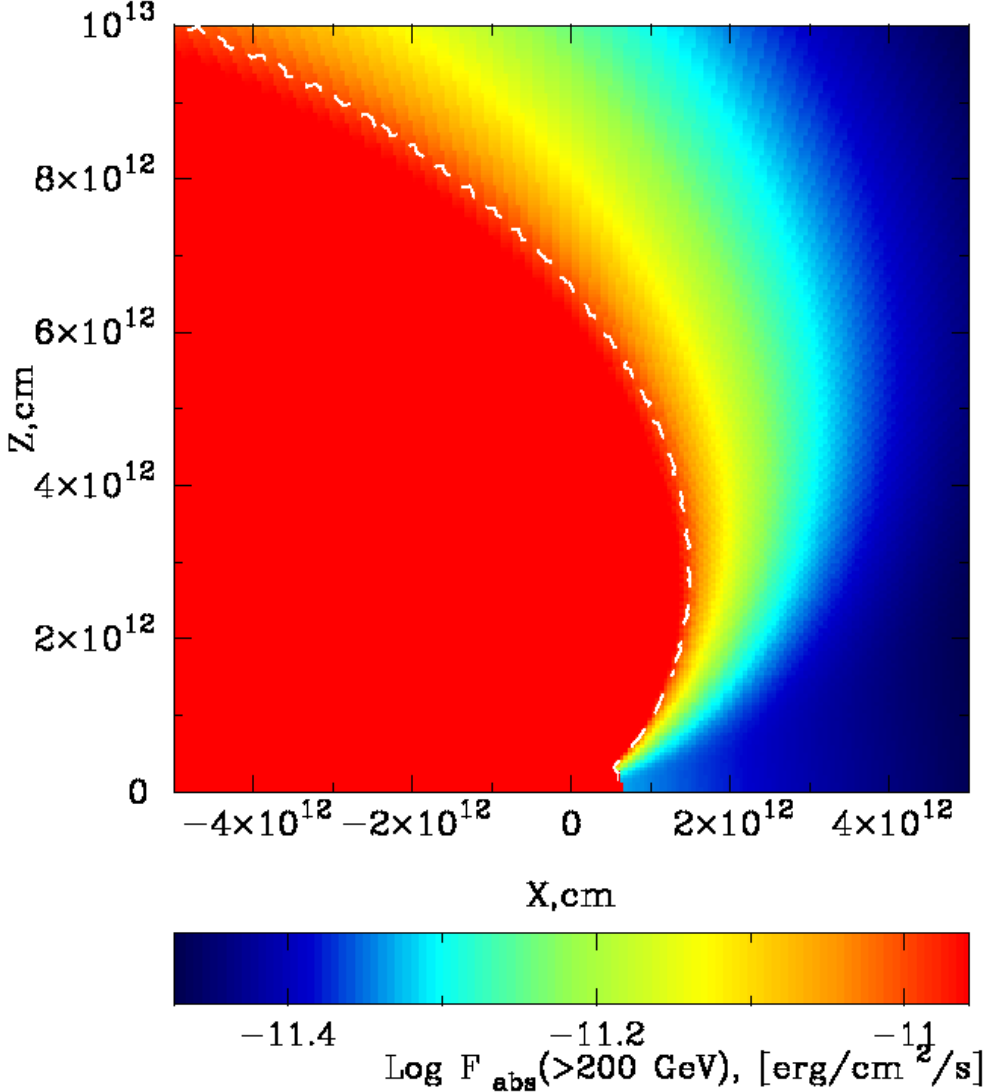}
\label{fig1}
\end{figure}

We have plotted a 2-dimensional map of the absorbed luminosity via photon-photon absorption  depending on the location of the
emitter in LS~5039. This is presented in Figure~\ref{fig1}. The XZ coordinates correspond to the plane formed by the emitter
and the star (0,0) positions, and the observer line of sight. To compute these maps, we have deabsorbed the observed spectra
and fluxes >100~GeV from LS~5039 around the superior conjunction of the compact object. The regions forbidden by the X-ray
observational constraints are limited by a contour line. The VHE emitter cannot be located to the left of the dashed line or
the X-ray observed fluxes will be overcome by the secondary pair synchrotron emission, expected to be the dominant secondary
pair cooling channel for reasonable magnetic field values in the stellar surroundings. O-star surface magnetic fields may
reach $\sim$~kG \cite{Donati et al. 2002}, and values of $\ge 10$~G could be realistic at few stellar radii from the stellar
surface, more than enough to suppress electromagnetic cascades \cite{Khangulyan et al. 2008}. In the plot, the compact object
is located in the left half of the XZ plane at $1.4\times 10^{12}$~cm from the star independently of the orbital inclination.

\begin{figure}
\caption{Computed spectral energy distribution of the synchrotron emission produced by the secondary pairs created in the
surroundings of the VHE emitter during the superior conjunction of the compact
object. The emitter location has been taken close to the compact object, with a
magnetic field of 10~G, and an inclination angle of $60^{\circ}$.}
\includegraphics[height=.4\textheight]{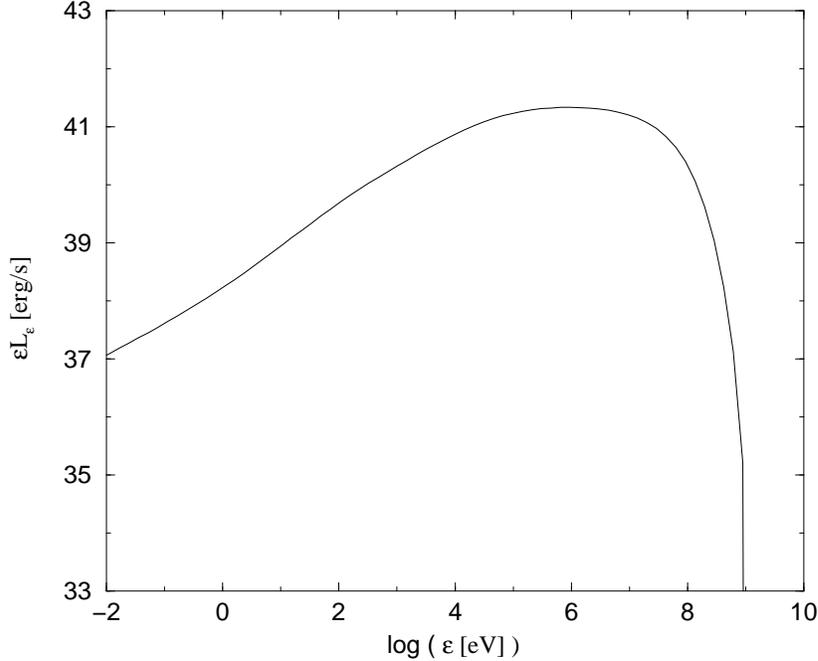}
\label{fig2}
\end{figure}

In Fig.~\ref{fig2}, we show the spectral energy distribution (SED) of the synchrotron emission produced by the secondary
pairs created in the surroundings of the VHE emitter during the superior conjunction of the compact object. The emitter
location has been taken close to the compact object, with an ambient magnetic field of 10~G, and an inclination angle of
$60^{\circ}$. This value for the inclination angle would correspond to the case of a neutron star as the compact object
\cite{Casares et al. 2005}, e.g. a pulsar. As seen in the figure, the resulting X-ray luminosity is seven orders of magnitude
larger than the one found by observations \cite{Bosch-Ramon et al. 2007}. As noted in Fig.~\ref{fig1}, only in case the
emitter were located far from the compact object, and far as well from the line joining the compact object and the star, the
synchrotron radiation would not overcome the observed fluxes (see also \cite{Bosch-Ramon et al. 2008}). 

Summarizing, the detection by HESS of VHE radiation from LS~5039 during the superior conjunction of the compact object, plus
a realistic ambient magnetic field, point strongly to an emitter located far away from the compact object, and far as well
from the region between the pulsar and the star. A jet-like emitter may play such a role, transporting energy to regions of
lower opacities; otherwise, the detected X-ray fluxes would be violated. This makes any scenario in which the TeV emission
would come from the region close to the compact object (the inner jet regions), or between the star and the compact object
(pulsar scenario for LS~5039; e.g. \cite{Dubus 2006b,Dubus et al. 2008,SierpowskaTorres2007,SierpowskaTorres2008}),
unlikely.  

\acknowledgments
V.B-R. gratefully acknowledges support from the Alexander von Humboldt
Foundation. V.B-R. acknowledges support by DGI of MEC under grant
AYA2007-68034-C03-01, as well as partial support by the European Regional
Development Fund (ERDF/FEDER).

\end{document}